\documentclass[%
 reprint,
superscriptaddress,
showpacs,preprintnumbers,
nofootinbib,
 amsmath,amssymb, 
 aps,
 prd,
 longbibliography,
]{revtex4-1}

\usepackage{cancel}
\usepackage{accents}
\usepackage{mciteplus,slashed}
\usepackage{amssymb,cancel,amsmath}
\usepackage{dcolumn}
\usepackage{bm}
\usepackage[caption=false]{subfig}
\usepackage{appendix}
\usepackage{physics}
\usepackage{booktabs}
\usepackage{feynmp-auto}
\usepackage[T1]{fontenc}	
\usepackage{csvsimple}
\usepackage{hyperref}
\usepackage[section]{placeins}
\usepackage[capitalise]{cleveref}
\usepackage{booktabs}
\usepackage{graphicx}
\usepackage{mathrsfs}
\graphicspath{{Figures/}}

\AtBeginDocument{
\heavyrulewidth=.08em
\lightrulewidth=.05em
\cmidrulewidth=.03em
\belowrulesep=.65ex
\belowbottomsep=0pt
\aboverulesep=.4ex
\abovetopsep=0pt
\cmidrulesep=\doublerulesep
\cmidrulekern=.5em
\defaultaddspace=.5em
}
\usepackage[dvipsnames]{xcolor}
\usepackage[normalem]{ulem}
\usepackage{fontawesome} 

\def\e{\mathrm{e}}
\def\Ldet{\ell}
\def\Ldec{\lambda}

\begin{document}

\title{Luminous solar neutrinos II: Mass-mixing portals}

\author{Ryan Plestid}
\email{rpl225@uky.edu}
\affiliation{Department of Physics and Astronomy, University of Kentucky  Lexington, KY 40506, USA}
\affiliation{Theoretical Physics Department, Fermilab, Batavia, IL 60510,USA}
\preprint{FERMILAB-PUB-20-522-T-V}

\begin{abstract}
     {\centering{\href{https://github.com/ryanplestid/luminous-solar-nu}{\large\color{BlueViolet}\faGithub}}  \\}  
     Solar neutrinos can be efficiently upscattered to MeV scale heavy neutral leptons (HNLs) within the Earth's mantle. HNLs can then decay to electron-positron pairs leading to energy deposition inside large-volume detectors. In this paper we consider mass-portal upscattering of solar neutrinos to HNLs of mass $20~\text{MeV}\geq m_N \geq 2 m_e$. The large volume of the Earth compensates for the long decay-length of the HNLs leading to observable rates of  $N\rightarrow \nu_\alpha e^+e^-$ in large volume detectors. We find that searches for mantle-upscattered HNLs can set novel limits on mixing with third generation leptons, $|U_{\tau N}|$ for masses in the MeV regime; sensitivity to mixing with first- and second-generation leptons is not competitive with existing search strategies.
\end{abstract}

 \maketitle 
 
 \section{Introduction \label{Introduction}}
\begin{figure}
    \subfloat[][]{\includegraphics[height=0.35\linewidth]{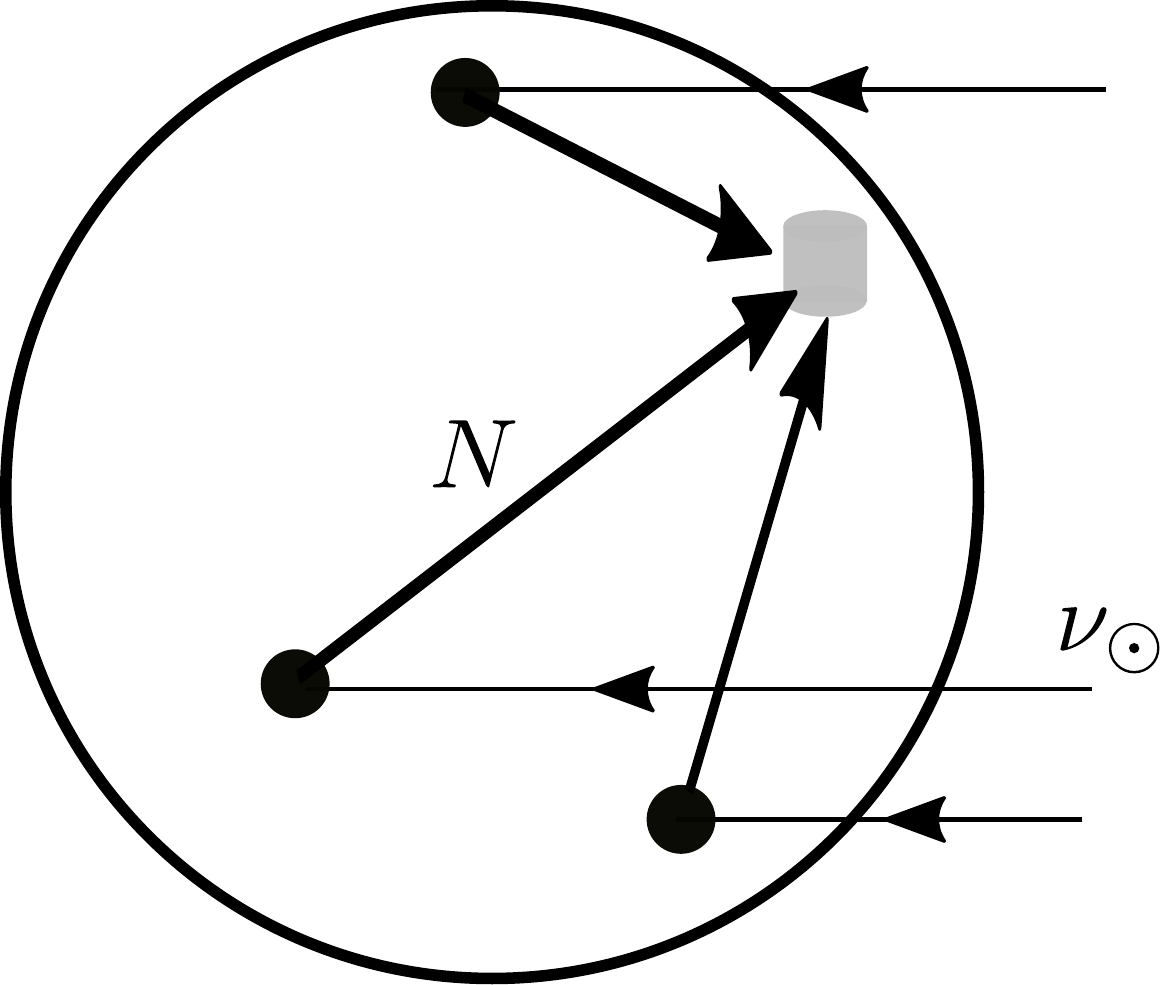}}%
    \qquad
    \subfloat[][]{\hspace{10pt} \includegraphics[height=0.35\linewidth]{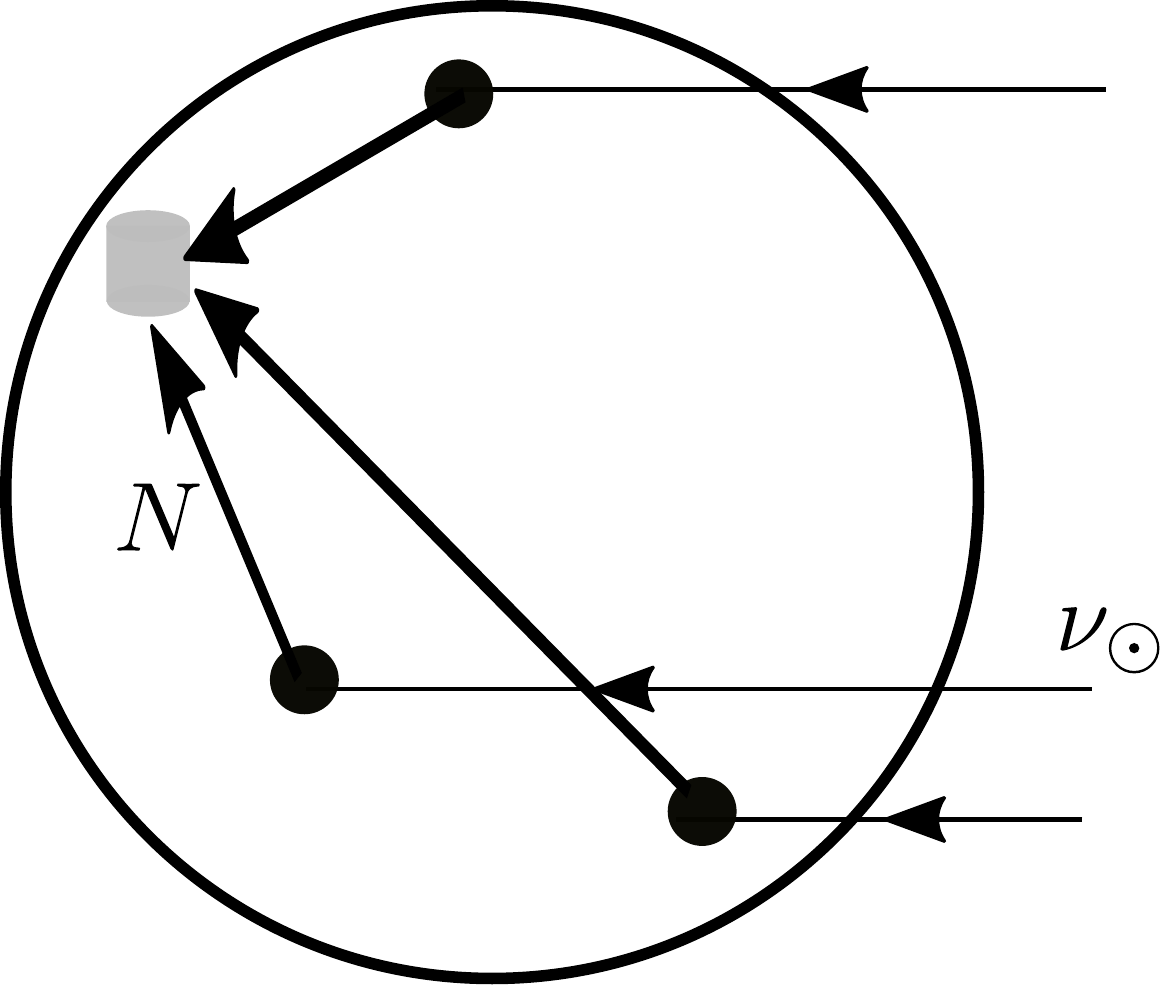}}
    \caption{Scatterings inside the volume of the earth that produce an HNL which subsequently decays to $e^+e^-$ inside a large volume detector (grey cylinder). Fig.\ (a) corresponds t0 day-time  (mostly backwards scattering) and Fig.\ (b) corresponds to night-time (mostly forward scattering). \label{Earth-sketch}}
\end{figure}

 Heavy neutral leptons (HNLs) are amongst the best motivated extensions of the Standard Model (SM) and their phenomenology has  been studied extensively for masses in the MeV to the few-GeV regime \cite{Arguelles:2019ziu,Liventsev:2013zz,Asaka:2005pn,Asaka:2005an,Atre:2009rg,Johnson:1997cj,Levy:2018dns,Formaggio:1998zn,Gorbunov:2007ak,Drewes:2013gca,Bondarenko:2018ptm,Boyarsky:2009ix,Berryman:2019dme,Ballett:2019bgd}. They have motivated dedicated detector technologies and are an ever-present driver of the physics case for fixed target facilities with high intensity beams \cite{Curtin:2018mvb,SHiP:2018xqw,Kling:2018wct,Hirsch:2020klk}. The lowest dimensional operator that mediates couplings to SM neutrinos is the so-called neutrino- or mass-mixing-portal \cite{Batell:2009di}. This leads flavor eigenstates, $\nu_a$,  to contain an admixture of HNLs, $N$, alongside the standard mass eigenstates, $\nu_i$,
 \begin{equation}
    \nu_a = U_{aN} N + \sum_{i=1}^3 U_{ai}\nu_i~.
 \end{equation}
 Constraints on $U_{\alpha N}$ in the literature stem mostly from Intensity Frontier experiments \cite{deGouvea:2015euy}, and are notably lacking at low masses\footnote{Bounds related to BBN rely on additional assumptions. Bounds from SN1987a rely on modelling of neutrino flavor composition in a supernovae.} where the decay length, $\Ldec$, of the HNL becomes very long. This is easy to understand since, in minimal HNL models with no additional degrees of freedom, the decay-rate of an HNL scales roughly as the muon-decay-like formula 
 \begin{equation}
    \Gamma\sim \frac{G_F^2 m_N^5}{192 \pi^3} ~\times \sum_{a\in \{e,\mu,\tau\} } |U_{aN}|^2~.
 \end{equation}
Thus, for small $m_N$,  the decay length, $\Ldec$, is enormous and any HNLs that are produced near a given experiment have a very small probability (of order $\Ldet/\Ldec$ with $\Ldet\sim 1-10$ m the length of the detector) of decaying within the volume of the detector. This can be circumvented with specialty built facilities such as SHiP \cite{SHiP:2018xqw}, FASER \cite{Feng:2017uoz,Kling:2018wct}, ANUBIS \cite{Hirsch:2020klk},  or MATHUSLA \cite{Curtin:2018mvb}, however this ultimately only extends the ``volume''  to $O(100~\text{m})$. In this paper we point out that for low masses (where decay lengths are thousands of kilometers or more), this problem can be circumvented by taking advantage of upscattering of solar neutrinos within the Earth's interior as depicted in \cref{Earth-sketch}. Similar ideas have been discussed in \cite{Bellini:2013uui} where the Borexino collaboration searched for HNLs that decay in flight after being produced in the Sun (a characteristic length scale of $10^{8}$ km), and in \cite{Arguelles:2019ziu} where mesons decaying in the upper atmosphere can produce HNLs that decay inside terrestrial detectors detectors (a characteristic length scale of 10 km).    %

 In a related paper \cite{Plestid:2020vqf}, we study a neutrino dipole portal \cite{Magill:2018jla,Brdar:2020quo} and show that the Earth's mantle can serve as a powerful resource when $\Ldec$ becomes very large ameliorating the naive $O(10~\text{m})/\Ldec$ suppression expected for a Borexino scale detector, and replacing it with the much more favourable $O(5000~\text{km})/\Ldec$ for $\Ldec \gg R_\oplus$.  This is a useful observation when coupled with the solar neutrino flux which contains sizeable $\nu_e$, $\nu_\mu$ and $\nu_\tau$ components.  Just like the neutrino dipole portal studied in \cite{Plestid:2020vqf}, upscattering of solar neutrinos via a mass-mixing portal is a coherent process (scaling as $Q_w^2$ with $Q_w$ the weak nuclear charge) in the $E_\nu \lesssim 20$ MeV regime allowing for further enhancement from the medium-heavy nuclei in the Earth's mantle. As we will show in this paper, the presence of a substantial $\nu_\tau$ flux allows us to set new constraints on mass-mixing portals connected to third generation leptons.

 There are, however, important differences between the phenomenology of mass-mixing and dipole-portal upscattering. Some differences are cosmetic, for instance the charge of the nucleus $Z$ being replaced by the weak charge $Q_w$, while others require a re-working of some of the formulae derived in \cite{Plestid:2020vqf}. For instance, the upscattering is now strongly energy dependent, scaling as $E_\nu^2$, and the decays are three-body. Most important, however, is that the scattering is no longer preferentially forward, being roughly isotropic in the lab frame. This is because the nuclei upon which the neutrinos scatter can be treated as being infinitely heavy, and the HNL that is produced can recoil in almost any direction. This eliminates most of the day-night asymmetry and seasonal signal modulation discussed in \cite{Plestid:2020vqf}, and demands an integration over the Earth's volume rather than the line of sight along the zenith direction. 
 
 The rest of this paper is dedicated to exploring the necessary modifications to the formalism developed in \cite{Plestid:2020vqf} to account for these differences. In \cref{Upscattering} we derive the flux of HNLs produced via mass-portal upscattering of solar neutrinos and discuss the different fluxes in the case of mono-flavor mixing (e.g.\ $U_{e4}=U_{\mu 4}=0$ with $U_{\tau 4} \neq 0$). In \cref{Decay} we discuss decay properties of $N$ and the resultant $e^+e^-$ rate inside a detector after accounting for scattering throughout the volume of the Earth.  In \cref{Constraints} we set limits on mixing matrix elements using Borexino's search for decaying HNLs from the Sun \cite{Bellini:2013uui}. Finally, in \cref{Conclusions} we summarize our results and comment on possible future improved sensitivity at future large-scale detectors.

\section{Mass-portal upscattering \label{Upscattering}}

For $\nu A\rightarrow N A$ scattering, the momentum transfer to the nucleus is limited by $2 E_\nu$. For solar neutrino energies this means that nucleus can be reliably approximated as an infinitely massive object. Within this approximation $E_N=E_\nu$ by energy conservation  and the matrix element for up-scattering is given, for a single mixing angle $U_{aN}$, by
\begin{equation}
    \left\langle|\mathcal{M}|^2\right\rangle= 8 M_A^2 |U_{aN}|^2 G_F^2 Q_w^2 \left(4 E_\nu^2-m_N^2+t\right).
\end{equation}
Using $\dd \sigma/\dd t  = \left\langle|\mathcal{M}|^2\right\rangle/(16\pi \lambda(s,M_A^2,m_\nu)$ with $m_\nu=0$ and $\lambda(a,b,c)$ the Kallen function, we have
\begin{equation}\begin{split}
    \dv{\sigma}{t}&=\frac{ |U_{aN}|^2 G_F^2 Q_w^2 E_\nu^2}{2\pi} \left(1-\frac{m_N^2}{4 E_\nu^2}+\frac{t}{4 E_\nu^2}\right).
\end{split}\end{equation}
At this point we can trade $t$ for the scattering angle of the HNL in the lab frame using 
\begin{equation}
    t= -|\vec{q}|^2= -\qty(E_\nu^2+ P_N^2 - 2 P_N E_\nu \cos\theta)~,
\end{equation}
where $P_N=\sqrt{E_\nu^2-m_N^2}$. For $m_N\ll E_\nu$ this gives 
\begin{equation}\begin{split}\label{costhetareal}
    \dv{\sigma}{\cos\theta}\propto 1+\qty[1-\frac{m_N^2}{2E_\nu^2}]\cos\theta .
\end{split}\end{equation}
Upscattering favors forward scattering, but only by a factor of roughly two. To simplify the analysis of upscattering within the earth we will approximate the differential upscattering cross section by 
\begin{equation}
    \dv{\sigma}{\Omega} = \frac{1}{4\pi} \sigma ~.
\end{equation}
This approximation underestimates the night-time rate (roughly by a factor of 50\%) and overestimates the day-time rate (roughly by a factor of 50\%) such that the full-day average is unaffected to a first approximation. 
\begin{figure}[b]
    \includegraphics[width=0.9\linewidth]{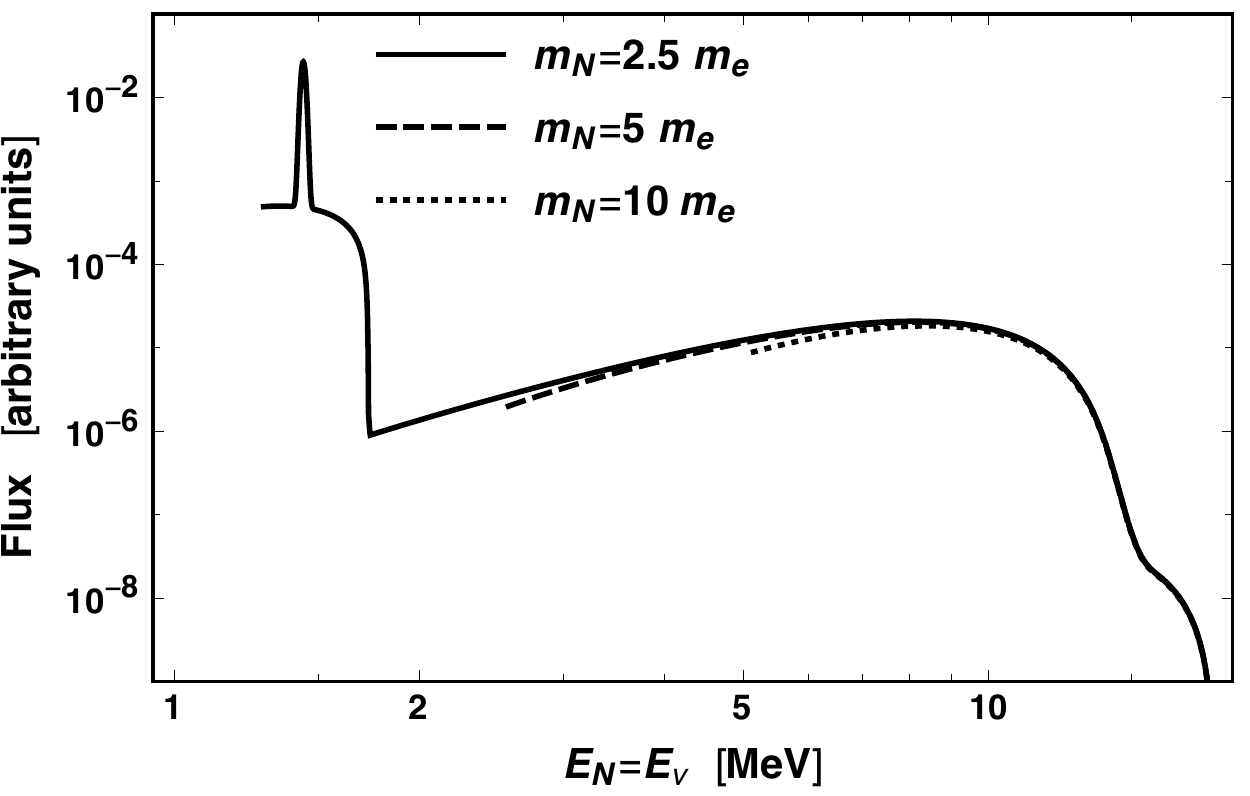}
    \caption{Shape of $N$ flux folded against $R_\oplus/\lambda$ for various choices of $m_N$ \emph{c.f.} \cref{Solar-Flux}. The flux of scattered $N$ is given by  $\Phi_N \propto  \Phi_{\nu}^{\odot} \times \sigma_{\nu\rightarrow N}$ and this is then multiplied by $1/\lambda(E_N)$. Here we have taken $\Phi_{\nu}^\odot =  \Phi_{\nu_\tau}^\odot$ as an example.  \label{N-spec}}
\end{figure}

The full cross section can be found by integrating $\dd \sigma/\dd t$ between the maximum and minimum momentum transfers
\begin{align}
    Q^2_\text{min} &= (P_\nu - P_N)^2 = (E_\nu- \sqrt{E_\nu^2 -m_N^2})^2 \\
    Q^2_\text{max} &= P_\nu^2 + P_N^2= 2 E_\nu^2 -m_N^2~. 
\end{align}
and clearly in the $E_\nu\gg m_N$ limit this give the result $Q_\text{min}= \tfrac12 m_N^2/E_\nu$ and $Q_\text{max}\approx 2E_\nu^2$. We then find
\begin{equation}\label{cross-section}
        \sigma_a= \frac{ |U_{aN}|^2 G_F^2 Q_w^2 E_\nu^2}{\pi} \left(3-\frac{m_N^2}{E_\nu^2}\right) \sqrt{1-\frac{m_N^2}{E_\nu^2}} ~.
\end{equation}
For a benchmark weak nuclear charge of $Q_w=1$ we then have
\begin{equation}\begin{split}
    \sigma_a= 4.7~\cdot ~&10^{-44}~\text{cm}^2 \qty[\frac{|U_{aN}|^2}{10^{-2}}]\qty[\frac{E_\nu^2}{10~\text{MeV}}]\\
    &\times \qty[\frac{3-m_N^2/E_\nu^2}{2.91}]\qty[\frac{1-m_N^2/E_\nu^2}{0.91}]^{1/2}
\end{split}\end{equation}  
The flux of HNLs, $\Phi_N$, emerging from an infinitesimal box of volume $\dd^3 x$, (shown in \cref{N-spec}) with target density $\overline{n}_A$, a distance $R$ away from the detector, is then given by 
\begin{equation}\label{flux-per-volume}
    \dd \Phi_N=    
    \frac{(\overline{n}_A \times \dd^3 x) \times \sigma_a \times \Phi^\odot_{\nu_a}\times \e^{-R/\Ldec}}{4\pi R^2} ~.
\end{equation}
where $E_N=E_\nu$ in the $M_A\rightarrow \infty$ limit and the exponential factor accounts for HNL decay.  For flavor dependent couplings oscillation effects must be included in the solar neutrino spectrum. We incorporate these effects assuming fully adiabatic flavor conversion \cite{Akhmedov:2004rq} using survival probabilities from \cite{Harnik:2012ni,Vedran-PC}. 

\section{Decays to electrons \& positrons  \label{Decay} } 

The decay of HNLs is extremely well studied in the literature \cite{Johnson:1997cj,Levy:2018dns,Formaggio:1998zn,Gorbunov:2007ak,Drewes:2013gca,Bondarenko:2018ptm}. For $2m_e\lesssim m_N\lesssim 100$ MeV the tree-level decay modes that are available to an HNL are $N\rightarrow 3\nu$ and $N\rightarrow \nu e^+e^-$. Below the threshold for electron-positron production, the only visible decay mode is the loop-mediated $N\rightarrow \nu \gamma$, but this is very small in minimal HNL models and we do not consider it here. 

For $U_{eN}\neq 0$ both charged- and neutral-currents can mediate the decay and these two contributions interfere with one another. For $U_{\mu N}\neq 0$ or $U_{\tau N} \neq 0$ only the neutral current participates. In both cases the partial width is given by \cite{Bondarenko:2018ptm}
\begin{widetext}
\begin{equation} \begin{split}
  \Gamma (N \to \nu_{a}e^+e^-) &= \frac{G_F^2 m_N^5}{192 \pi^3}                                      
  \cdot |U_{aN}|^2 \cdot \bigg\{ C_1 \qty[(1 - 14 x^2 - 2 x^4 - 12 x^6)                                   
  \sqrt{1 - 4 x^2} + 12 x^4 (x^4 - 1)L(x)]\\
  &\hspace{115pt}+ 4 C_2 \qty[x^2 (2 + 10 x^2 - 12 x^4)\sqrt{1 - 4 x^2} + 6 x^4 (1 - 2 x^2 + 2 x^4) L(x)] \bigg\},  \label{full-decay}
\end{split}\end{equation}
\end{widetext}
where $x=m_e/m_N$, and 
\begin{equation}
    L(x) = \log \bigg[ \dfrac{1 - 3 x^2 - (1 - x^2) \sqrt{1 - 4 x^2}}{x^2(1 + \sqrt{1 - 4 x^2})} \bigg]~. 
\end{equation}
The coefficients $C_1$ and $C_2$ are given in terms of the Weinberg angle as
\begin{align}
    C_1 &= \frac{1}{4}\bigg(1 \pm 4\sin^2 \theta_W + 8\sin^4\theta_W \bigg) \label{c1}\\
    C_2 &= \frac{1}{2} \sin^2 \theta_W \bigg(2 \sin^2 \theta_W \pm 1 \bigg)~,\label{c2}
\end{align}
with the upper signs (+) corresponding to $a=\mu$ and $a=\tau$, and the lower (-) signs corresponding to $a=e$. 

The probability of an HNL decaying to $e^+e^-$ in some distance $\ell\ll\Ldec$ is given by $\text{BR}(e^+e^-)\ell/\Ldec$. In the limit when $\Ldec \gg R_\oplus$ the decay length will always appear as $\text{BR}/\Ldec= \Ldec_{e^+e^-}$ which is the decay length one would find if only the $e^+e^-$ decay pathway were considered. Since, for minimal models of HNLs without an augmented dark sector, the condition $\Ldec \gg R_\oplus$ is always satisfied, we ignore invisible decay modes hereafter and use $\lambda$ and $\lambda_{e^+e^-}$ interchangeably. For non-minimal scenarios that lead to $\Ldec \lesssim R_\oplus$ (as in  e.g.\ \cite{Ballett:2019pyw,Bertuzzo:2018itn,DeRomeri:2019kic,HostertLOI} and as is discussed briefly in \cref{non-minimal}) one must include the invisible decay modes explicitly. 

The decay length of $N$ is then given by
\begin{equation}
\begin{split}\label{Ldec-benchmark}
    \Ldec &= \frac{E_N}{m_N} \sqrt{1-\frac{m_N^2}{E_\nu^2}} \frac{1}{\Gamma_{N}}\\
    &\approx  3.1\cdot10^{6} R_\oplus~~ \qty[ \frac{10^{-2}}{|U_{aN}|^2}]\qty[\frac{3~\mathrm{MeV}}{m_N}]^6\\
    &\hspace{45pt}\times \qty[\frac{E_N}{10~\mathrm{MeV}}]\sqrt{\frac{1-m_N^2/E_N^2}{0.91}} 
\end{split}
\end{equation}
where the approximation holds in the $x\rightarrow 0$ limit of \cref{full-decay}. We have assumed that $a\neq e$ such that we take the upper signs in \cref{c1,c2}. Clearly for any reasonable parameter choices relevant for solar neutrino upscattering we are in the limit where $\Ldec\gg R_\oplus$. 

Once the flux of $N$'s has been calculated the resultant spectral shape of positrons and electrons from $N\rightarrow \nu_a e^+e^-$ could be obtained from a first principles calculation. A proper treatment of the resultant signal in a detector is relatively involved since the signature will depend on the energy of the electron, the energy of the positron, the opening angle between them and detector details. The energy spectrum itself will depend on whether the HNL is Dirac or Majorana \cite{Berryman:2019dme}, and on the mass-dependent polarization inherited from the HNL's nascent production in a neutrino upscattering event (see e.g.\ \cite{Johnson:1997cj,Levy:2018dns,Formaggio:1998zn}). We leave these details to future work and focus our attention on a rate only analysis. 

The rate of deposition of $e^+e^-$ pairs inside a detector from an infinitesimal volume element, $\dd^3 x$, a distance $R$ from the detector is given by 
\begin{equation}\begin{split}
    \dv[3]{R}{x} = \int \dd E_N&~ (1-\e^{-\Ldet/\Ldec})~\e^{-R/\Ldec} \dv[3]{\Phi_N}{x} .
\end{split}\end{equation}
The flux per-unit volume is given in \cref{flux-per-volume}. We will always have $\Ldet \ll \Ldec$ such that $1-\e^{-\Ldet/\Ldec}\rightarrow \Ldet/\Ldec$. Multiplying by the detector's cross sectional area, $A_\perp$, and integrating over $x$ we have 
\begin{widetext}
\begin{equation}\label{Earth-Rate}
  R_{e^+e^-} = V_\text{det} 
    \int_{m_N}^{18.8~\text{MeV}} \dd E_N~\int_\oplus \dd^3 x ~\overline{n}_A (x) \frac{1}{4\pi R^2}\qty(\frac{\e^{-R/\Ldec}}{\Ldec}) \Phi^\odot_{\nu_a} (E_\nu = E_N)  \sigma_a (E_N)
\end{equation}
\end{widetext}
with $V_\text{det}= A_\perp \Ldet$ and the solar neutrino flux $\Phi_{\nu_a}^\odot$ evaluated at $E_\nu=E_N$. We have implicitly assumed that $\Phi_{\nu}^{\odot}(E_\nu)$ is independent of the position,  $x$, within the mantle, which is a good approximation for solar neutrinos.

In the minimal models of HNLs that we focus on here, the decay length is always much much greater than the radius of the Earth. Taking the density inside the Earth to be constant, $\overline{n}_A(x)=\overline{n}_A$, the volume integral in \cref{Earth-Rate} simplifies to 
\begin{equation}\label{main-rate-0}
   R_{e^+e^-} = [V_\text{det} \overline{n}_A]
    \int_{m_N}\hspace{-6pt} \dd E_N \frac{R_\oplus}{2 \Ldec}~\Phi^\odot_{\nu_a}  \sigma_a~.
\end{equation}
The decay length, $\Ldec$, neutrino flux, and upscattering cross section are all functions of $E_\nu=E_N$. 

\Cref{main-rate} has been derived assuming a constant density profile throughout the entirety of the Earth, however, as we discuss in \cref{Earth-Material} it can be easily modified to account for the added flux coming from the high density core of the Earth. Including the Earth's core enhances the flux of HNLs arriving at the detector by a factor of 2.44 (see \cref{eff-nAQw} and discussion thereafter). We therefore have
\begin{equation}\label{main-rate}
   R_{e^+e^-} =2.44 [V_\text{det} \overline{n}_A Q_w^2]_\text{mantle}
    \int_{m_N}\hspace{-6pt} \dd E_N \frac{R_\oplus}{2 \Ldec}~\Phi^\odot_{\nu_a}  \frac{\sigma_a}{\overline{Q}_w^2}~.
\end{equation}
This is the main equation we will make use of in setting limits. We take $\overline{Q}_w=12.2$ and $\overline{n}_A=0.95\cdot 10^{23}$ cm$^{-3}$.

\section{New constraints on mass-mixing  \label{Constraints}}
The spectrum of $e^+e^-$ pairs will lie between roughly a few MeV and the maximum energy allowed by the solar neutrino flux, roughly 18.8 MeV. A comprehensive treatment would include the anisotropy of the upscattering,  the full spectral shape of the $e^+$ and $e^-$, and the correlations between $e^+$ and $e^-$ pairs. This would lead to the following phenomenological signatures
\begin{itemize}
    \item The directionality of the $e^+e^-$ pair will not have a strong correlation with the position of the Sun. 
    \item The signal will modulate by an $O(1)$ factor between day and night because \cref{costhetareal} prefers forward scattering by an $O(1)$ amount. 
    \item The $e^+e^-$ pair will be coincident in time.
    \item Spectral shape information can be used to suppress backgrounds. This requires an understanding of the HNL polarization and its Dirac vs Majoran nature.
\end{itemize}
Such an analysis lies well beyond the scope of this work and we focus here, instead, on a simple rate-only estimate. We focus on the time averaged rate (justifying our treatment of the scattering as isotropic) taken over a full year using data from Borexino to set constraints.

We use the collaboration's search for $N\rightarrow e^+e^-\nu$ from HNLs produced in the Sun that decay in flight.  Just like in the decay in flight search, for our upscattering scenario the $e^+e^-$ events from $N$ decay  will be uniformly distributed throughout the volume of the detector and we therefore expect the signal yield to be negligibly affected by the collaboration's cuts on their data in both cases. We therefore consider the observed event rate after cuts have been imposed, focusing on the total number of events in 446 days of live time as presented in Fig.\ 4 of \cite{Bellini:2013uui}. We see that the collaboration measured roughly 75 events from 4.8 MeV to 12.8 MeV and that the expected solar neutrino background from $^{8}$B neutrinos was roughly 65 events. The collaboration expected no other backgrounds but allowed for $^{11}$Be decays as a potential source of contamination. In setting their limits on $|U_{eN}|$ the collaboration allowed the number of $^{11}$Be events and the $^{8}$B spectrum to float.  In the same figure the expected flux from HNLs produced in the Sun that decay in flight is shown for $|U_{eN}|^2=8\times 10^{-6}$. The shown rate corresponds to roughly 15 total events which the collaboration later (comfortably) excludes on the basis of a statistical treatment (see Fig.\ 5 \& 6 of the same paper). We therefore take 15 events in 446 days of runtime with a 100 tonne fiducial volume as a benchmark rate for limit setting purposes.

To build some intuition about the size of the rates that we are considering, it is convenient to re-express \cref{main-rate} in terms of benchmark parameters 
\begin{widetext}
\begin{equation} \label{main-rate-explicit}
R_{e^+e^-}= 0.17 \text{~yr}^{-1} \bigg[\frac{V_\text{det} \overline{n}_A}{10^{31}}\bigg] \bigg[\frac{\overline{Q}_w}{12}\bigg]^2 \bigg[\frac{m_N}{3~\text{MeV}}\bigg]^6\bigg[ \frac{|U_{aN}|^4}{10^{-4}}\bigg] \int_{m_N}^{18.8~\text{MeV}}\hspace{-15pt}\dd E_N \bigg[\frac{3-m_N^2/E_N^2}{2.91}\bigg] \bigg[\frac{E_N}{10~\text{MeV}} \bigg] \bigg[\frac{\Phi^\odot_{\nu_{a}}(E_N)}{10^8~\text{cm}^2~\text{s}^{-1}}\bigg]~,
\end{equation}
\end{widetext}
which is valid in the $x\rightarrow 0$ limit where $x=m_e/m_N$. In setting our limits we include the full $x$ dependence of \cref{full-decay} as a prefactor multiplying \cref{main-rate-explicit}. This suppresses the decay close to the $e^+e^-$ threshold, and becomes an  $O(1)$ prefactor for $m_N\gtrsim 2$ MeV. 

\begin{figure}
    \includegraphics[width=\linewidth]{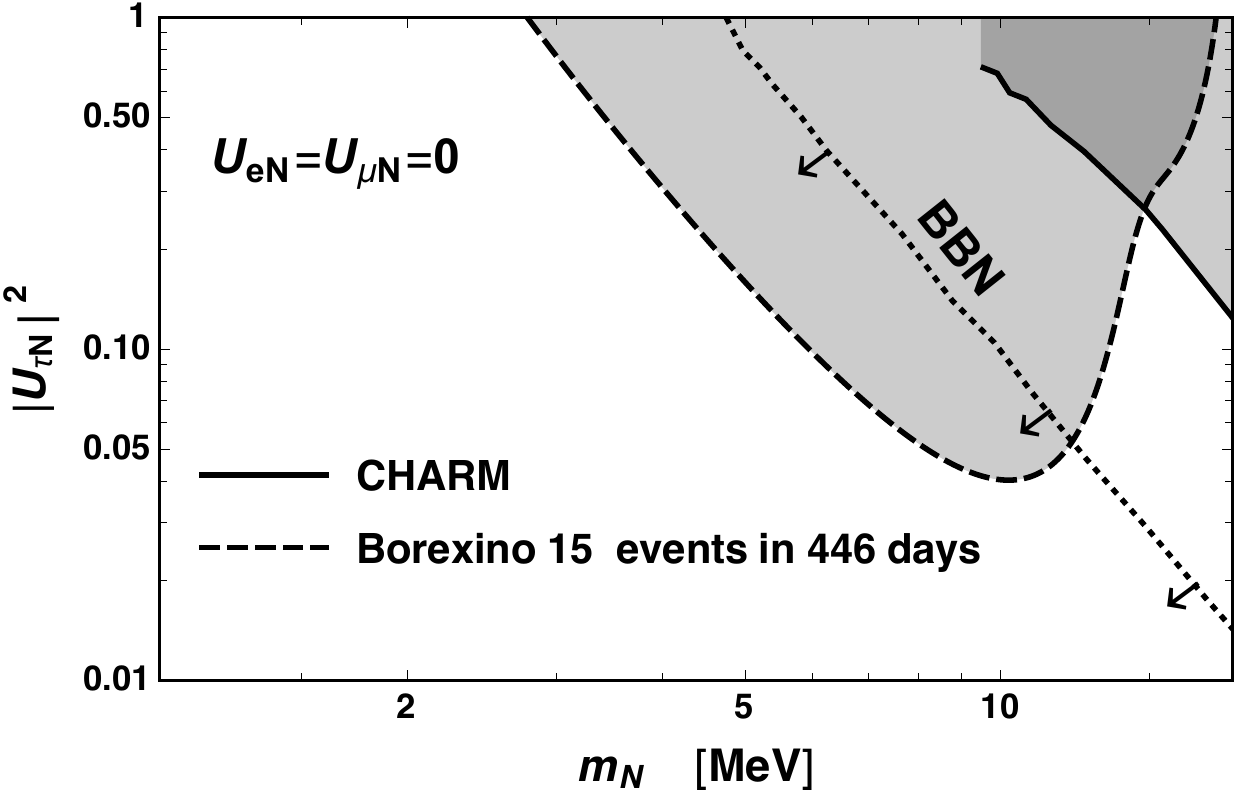}
    \caption{Parameter space that would lead to greater than 15 expected events in the Borexino 100 tonne fiducial volume in 446 days of run time.  Constraints from the CHARM collaboration \cite{Orloff:2002de} are shown as a solid line. Constraints from BBN \cite{Sabti:2020yrt} are shown as a dotted line, with the region being pointed to by the black arrows being excluded.  One sees that solar neutrino upscattering complements bounds from cosmology and the CHARM collaboration.  \label{tau-sensitivity}}
\end{figure}

The rates are very small and for allowed values of mixing with first- and second-generation leptons, they yield signals that are entirely unobservable. Mixing with third generation leptons, i.e.\ $U_{eN}=U_{\mu N}=0$ and $U_{\tau N}\neq 0$,  is almost completely devoid of existing constraints at low HNL masses. The only notable constraint comes from the CHARM collaboration \cite{Orloff:2002de}, and below $m_N \lesssim 10$ MeV there are no constraints at all. The solar neutrino flux contains a sizeable $\nu_\tau$ component, and so we find that upscattered solar neutrinos can provide novel constraints on (and possibly lead to future discoveries of) mass-mixing with third generation leptons. We show our results  in \cref{tau-sensitivity}.

Given that solar neutrino upscattering naturally probes parameter space with very long decay lifetimes, it is interesting to ask whether the parameter space of interest is compatible with cosmological bounds. Surprisingly, because the mixing angle that Borexino is able to probe are relatively sizeable, lifetimes are sufficiently short such that Borexino is able to provide novel constraints on parameter space that is untouched by either existing data from accelerator experiments or analyses of big bang nucleosynthesis (BBN). To see this explicitly we have included the bounds derived in Fig.\ 10 of \cite{Sabti:2020yrt} for comparison in \cref{tau-sensitivity}. 

\section{Conclusions \label{Conclusions}} 

The volume of the Earth can serve as a powerful resource for upscattering astrophysical particles that, if they are long lived but unstable, can leave visible imprints in large volume detectors. In this work we have shown that even in regimes where signal rates are low (because of prohibitively long decay lengths) new insights on well motivated BSM models can be gleaned. The presence of (otherwise hard to come by) tau-neutrinos makes even the modest signal yields from $\nu_\tau \rightarrow N$ a valuable tool in the study of HNL physics.  

The reason that the signal yields turned out to be so small is primarily due to the fact that at low energies HNLs have enormously long decay lengths, six orders of magnitude larger than the radius of the Earth. These distances are comparable to the distance between the Earth and the Sun, and this fact is what allowed Borexino to obtain strong constraints on $\nu_e-N$ mixing \cite{Bellini:2013uui}. Because the neutrinos participating in nuclear reactions inside the sun are electron flavor eigenstates, no such constraints could be placed on $\nu_\tau-N$ mixing by considering decay in flight decays of HNLs produced in the Sun. In contrast we make use the oscillated  flux arriving at earth which has $\nu_e$, $\nu_\mu$ and $\nu_\tau$ components. 

Much larger signal yields can be easily imagined in models with an interacting dark sector. For example, if the $N$ couples to some $Z'$, then $N\rightarrow Z' \nu$ followed by $Z'\rightarrow e^+e^-$ could yield a decay length $\Ldec' $ that is comparable to or smaller than the size of the Earth. In such scenarios, there is no $R_\oplus/\lambda$ suppression and the signal would scale as $|U_{aN}|^2$ rather than $|U_{aN}|^4$; the rate would also cease to carry a steep $m_N^6$ dependence on the mass of the HNL. Investigations into, and motivations of, such non-minimal scenarios are beyond the scope of this work, however these basic qualitative ideas should be kept in mind for the interested phenomenologist.  

Searching for signals of new physics emanating out of the Earth's mantle is a relatively unexplored idea. A literature exists on luminous dark matter \cite{Feldstein:2010su,Pospelov:2013nea,Serenelli:2011py,Eby:2019mgs} and this paper's companion \cite{Plestid:2020vqf} discusses neutrino dipole portals. We have so far, however, constrained ourselves only to the solar neutrino flux. Higher energy components of the astrophysical neutrino flux may be relevant in higher mass regions of the HNL parameter space where light-meson constraints, such as those stemming from $\pi \rightarrow e\nu$, become inapplicable \cite{deGouvea:2015euy}. A broadening physics case for such signals of ``luminous'' physics emanating from the interior of the Earth further motivates dedicated searches at large volume detectors, and a comprehensive experimental search  strategy.

\section{Acknowledgements}
  I would like to thank Joachim Kopp, Vedran Brdar, Kevin Kelly, Pedro Machado, Gordan Krnjaic, and Volodymyr Takhistov for helpful discussions. I would especially like to thank Matheus Hostert for pointing out Ref.\ \cite{Bellini:2013uui} to me during the Snowmass Mini Workshop on Neutrino Theory and and Patrick Fox for helpful feedback on early versions of this manuscript. I would like to thank the Fermilab theory group for their hospitality and welcoming research atmosphere.  My visit at Fermilab was supported with funds from the Intensity Frontier Fellowship. This work was supported by the U.S. Department of Energy, Office of Science, Office of High Energy Physics, under Award Number DE-SC0019095. This manuscript has been authored by Fermi Research Alliance, LLC under Contract No. DE-AC02-07CH11359 with the U.S. Department of Energy, Office of Science, Office of High Energy Physics. 
 
\appendix

\section{Effective composition of the Earth \label{Earth-Material}}

\begin{table}
\centering
$\begin{array}{ cccc}
\text{Molecule.} 		& n [10^{22}~ \text{cm}^{-3}]  & \Sigma N^2 & n \times \qty[ \Sigma N^2 ]  [10^{24} ~\text{cm}^{-3}]  	 \\
\midrule

\text{SiO}_2 & 1.81 &  324 & 5.85 \\ 
\text{MgO }& 2.33 &  208 & 4.85 \\ 
\text{FeO} & 0.28 &  964 & 2.66\\ 
\text{CaO} & 0.14 &  464  & 0.64 \\ 
\text{Cr}_2\text{O}_3  & 0.01 & 1760 &  0.16\\
\text{Ni}\text{O} & 0.01 &  1025 & 0.08 \\ 
\midrule
\text{Eff.\ Avg.\ } & 0.951 & 150.8 & 1.43\\
\midrule
\midrule
\text{Fe}  & 12.05 & 900 &  100\\
\text{O} & 3.91 &  64 & 2.48 \\ 
\midrule
\text{Eff.\ Avg.\  } & 13.9 & 795.2 & 111 \\
 \bottomrule
\end{array}$

\caption{Above double-rule: Mantle composition calculated using Table 3.\ of  \cite{WORKMAN200553} and a density of $\overline{\rho}=4$ g/cm$^3$. Elements with $n_i \leq 7 \cdot 10^{19}~\text{cm}^{-3}$ are omitted. Below double-rule: Core composition assuming the core is 90\% iron and 10\% oxygen with a density of $\overline{\rho}=13.5\text{g/cm}^3$. In all cases the neutron number is calculated using each element's most commonly occurring isotope. 
\label{earth-table}} 
\end{table}

In \cite{Plestid:2020vqf} the flux of solar neutrinos only encounters the core of the Earth  at certain times of the year. In contrast, a mass-mixing portal's isotropic upscattering and very long decay lengths ensures that the Earth's core always contributes to the upscattering. Unlike in \cite{Plestid:2020vqf}, we therefore include the core of the Earth in our discussion. 

 As a simple model of the Earth's composition we assume the core to be composed of a uniform density sphere of radius $R_\text{core}=0.5 R_\oplus$, and the rest of its volume to composed of uniform density mantle. We treat the core as having a uniform density of $13$ g/cm$^3$ and a composition that is 90\% $^{56}$Fe by mass and 10\% $^{16}$O by mass.

For the mantle we take a density of $\overline{\rho}=4$ g/cm$^3$. The elemental composition of the Earth's mantle is given in Table 3.\ of \cite{WORKMAN200553}. We use the DMM column, which gives the mass percentage by molecular compound  and the numerical values are summarized in \cref{earth-table}.  We are ultimately interested in 
\begin{equation}
    \overline{n}_A \overline{Q}_w^2 = \sum_{i} n_i \qty[\sum Q_w^2]_i 
\end{equation}
where the bracketed sum is adding up $Q_w^2$ within the atoms of each molecular compound. The quantity $Q_w$ is the charge of the weak charge of the nucleus. We neglect the proton contribution due to its small charge $1-4\sin^2 \theta_w\approx 0$. We therefore approximate $Q_w\approx N$.

The number densities can be re-expressed in terms of the mantle density via
\begin{equation}
    n_i = \frac{\overline{\rho}}{m_i} f_m 
\end{equation}
where $f_m$ is the mass fraction given in Table 3. of \cite{WORKMAN200553} (Bulk DMM). We can define $\overline{Q}_w^2$ via 
\begin{equation} 
   \overline{Q}_w := \frac{\langle N^2 \rangle }{\langle N \rangle} ~, 
\end{equation}
which is a density independent definition. The effective number density is then defined as 
\begin{equation}
   \overline{n}_A= \frac{ \sum_{i} n_i \qty[\sum N^2]_i }{\overline{Q}_w^2} ~.
\end{equation}
The same basic formulae apply also to the core. 

To incorporate the core and mantle together we can consider the following integrals 
\begin{align}
    I_1&=  [\overline{n}_A \overline{Q}_w^2]_\text{core} \int_\text{core} \dd^3 x~ \frac{\tfrac1\lambda\e^{-R/\lambda}}{4\pi R^2} \\
    I_2 &= [\overline{n}_A \overline{Q}_w^2]_\text{mantle} \int_\text{mantle} \dd^3 x~ \frac{\tfrac1\lambda\e^{-R/\lambda}}{4\pi R^2}
\end{align}
This then defines an effective value of $\overline{n}_A \overline{Q}_w^2$ via 
\begin{equation}\label{eff-nAQw}
    [\overline{n}_A \overline{Q}_w^2]_\text{eff} = \frac{I_1 + I_2}{\tfrac1\lambda\int_\oplus \dd^3x~\frac{\e^{-R/\lambda}}{4\pi R^2}}
\end{equation}
We find that for the composition model in \cref{earth-table} that $[\overline{n}_A \overline{Q}_w^2]_\text{eff} = 2.44 [\overline{n}_A \overline{Q}_w^2]_\text{mantle}$ and we include this enhancement from the core in our results.

\section{Solar Neutrino Flux}

\begin{figure}[hb]
    \includegraphics[width=0.9\linewidth]{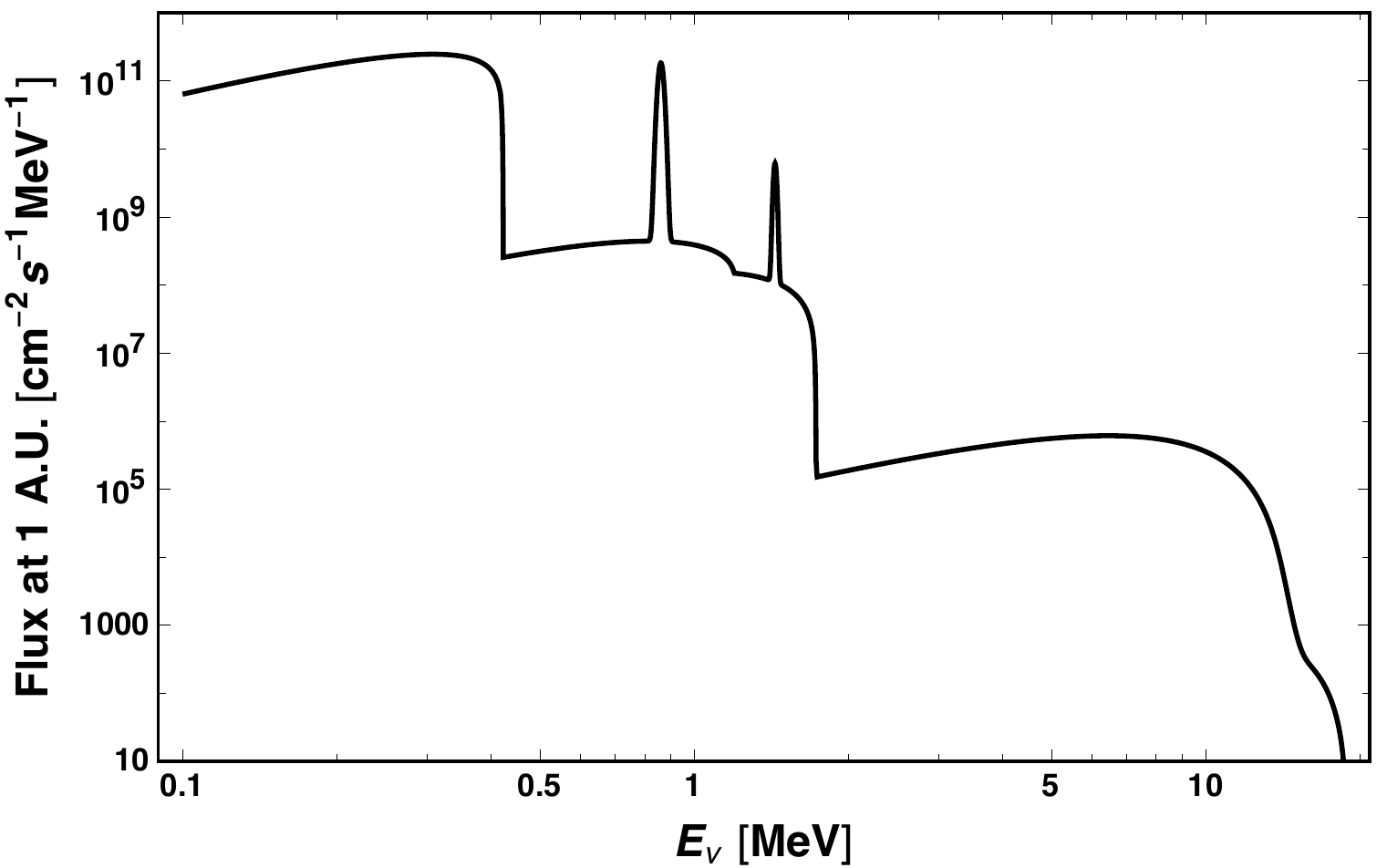}
    \caption{Solar neutrino flux.  Shapes are taken from \cite{Bahchall_url} and normalizations from Tab.\ 2 of \cite{Serenelli:2011py} (AGSS09 \cite{Asplund:2009fu}).\label{Solar-Flux} }
\end{figure}

For completeness we describe our input solar neutrino flux. This is most easily summarized in visual form in \cref{Solar-Flux}. The $^7$Be line at 384 keV has been included but is sub-dominant to the pp and hence not visible. We treat the. $pep$ and $^7$Be lines as Gaussians with a width of 10 keV.

\section{Shorter decay lengths \label{non-minimal}}

In non-minimal models (see e.g.\ \cite{Ballett:2019pyw,Bertuzzo:2018itn,DeRomeri:2019kic,HostertLOI}) of HNLs where there are substantial branching ratios to short-lived $Z'$s (for example), which then decay primarily into $e^+e^-$ pairs, the decay length may be comparable to, or shorter than, the radius of the Earth. In this case a proper accounting for the integral \cref{Earth-Rate} is needed. In these cases, for a surface level detector we find
\begin{equation}
    \int_\oplus \dd^3 x ~\frac{\tfrac1\lambda\e^{-R/\lambda}}{4\pi R^2} = \frac{\Ldec}{4 R_\oplus} \qty[ e^{-2 R_\oplus/\Ldec}-\qty(1-\frac{2R_\oplus}{\Ldec})]~, 
\end{equation}
with $R=|\vb{x}_\text{det}-\vb{x}|$. For $\Ldec \ll R_\oplus$ we find that the integral is \emph{independent} of $\Ldec$ because the small probability of decaying inside the detector is compensated for by a larger number of upscattering targets lying within one decay length of the detector.  In these scenarios, the proper decay length must be used and the branching factor accounted for since the assumptions outlined in footnote 2 are broken.

\vfill

\bibliography{biblio.bib}
\end{document}